\documentclass[pra,aps,groupaddress,showpacs]{revtex4}
\usepackage{epsfig,amsmath,graphicx,amssymb,overpic}
\usepackage{bm}
\usepackage{mathrsfs}
\usepackage{graphicx}
\usepackage{epsfig}
\usepackage{dcolumn}
\usepackage{bm}
\usepackage{amsmath,amssymb,amsthm}
\usepackage[colorlinks=true,linkcolor=blue]{hyperref}
\usepackage{subfigure}
\usepackage{booktabs}
\usepackage[mathscr]{euscript}
\usepackage{ulem}

\newcommand{\bea}{\begin{eqnarray}}
\newcommand{\eea}{\end{eqnarray}}
\newcommand{\beq}{\begin{equation}}
\newcommand{\eeq}{\end{equation}}

\def\/{\over}

\usepackage{CJK}
\begin{document}
\title{Decomposition solutions and B\"acklund transformations of the BKP and CKP equations}
\author{Xiazhi Hao$^1$ and S. Y. Lou$^2$\thanks{Corresponding author:lousenyue@nbu.edu.cn. Data Availability Statement: The data that support the findings of this study are available from the corresponding author
upon reasonable request.}}
\thanks{Corresponding author: lousenyue@nbu.edu.cn. Data Availability Statement: The data that support the findings of this study are available from the corresponding author
	upon reasonable request.}
\affiliation{\footnotesize{$^1$College of Science, Zhejiang University of Technology, Hangzhou, 310014, China}\\
\footnotesize{$^2$School of Physical Science and Technology, Ningbo University, Ningbo, 315211, China}}

\begin{abstract}
	In this paper, we define the modified formal variable separation approach and show how it determines, in a remarkably simple manner, the decomposition solutions, the B\"acklund transformations, the Lax pair, and the linear superposition solution of the B-type Kadomtsev-Petviashvili equation. Also, the decomposition solutions, the B\"acklund transformation and the Lax pair relating to the C-type Kadomtsev-Petviashvili equation is obtain by the same technique. This indicates that the decomposition may provide a description of integrable behavior in nonlinear systems, while, at the same time, establishing an efficient method for determining relationships between the particular systems.
\\
{\bf Key words: \rm Decomposition solution, B\"acklund transformation, Lax pair, Linear superposition solution}
\end{abstract}

\pacs{05.45.Yv,02.30.Ik,47.20.Ky,52.35.Mw,52.35.Sb}
\maketitle

\section{Introduction}

Various techniques are available for obtaining exact solutions of nonlinear partial differential equations (PDEs) such as the inverse scattering method \cite{abib14,rbib22,vbib4}, group theoretical method \cite{abib23}, the singularity approach \cite{jbib3,jbib9,rbib3,sbib8}, and some direct methods \cite{xbib1,ebib14,ybib4,wbib25}. In all these methods, one can identify a central link, B\"acklund transformation (BT for short), one analytical tool for dealing with integrability problems, which enables one to relate pairs of solutions of certain nonlinear PDEs. 
Indeed the number of papers produced so far on BTs is fairly large, but not surprisingly so since these transformations have attracted the attention of both mathematicians and physicists, and have been incredibly actively developed in different directions, their relevance being well established in differential geometry and algebra as well as in nonlinear science, and other fields of application \cite{gbib16,mbib2,mbib15,jbib10}.

Nevertheless, among these methods, it seems to us that the convenient one to construct BTs is the decomposition method based on the formal variable separation approach (FVSA) \cite{sbib42,sbib43,xbib26}. We propose such a new decomposition through which the BTs may exist and show how it works for two important nonlinear PDEs: the B-type and C-type Kadomtsev-Petviashvili equations (BKP and CKP). In the authors' opinion, this decomposition provides a very straightforward and natural way to construct BTs, in addition, yields some new results. It may be expected that these results can be carried over to other equations.

The modified FVSA consists in looking for the general decomposition solution of the (2+1)-dimensional PDE in the form 
	\begin{eqnarray}\label{gwy}
	w_y& =& F(x, y, t, w, w_x, w_{xx},\dots, w_{xm}), \\\label{gwt}
	w_t &=& G(x, y, t, w, w_x, w_{xx},\dots, w_{xn})
	\end{eqnarray} 
	with the integrability condition 
	\begin{eqnarray}\label{ic}
	w_{yt}-w_{ty}=[F,G]=0
	\end{eqnarray}
	preserved, here, $w_{xm}=\partial^m_xw, w_{xn}=\partial^n_xw,$ $m$ and $n$ are integers. $F$ and $G$ depend on $x,y,t$ and derivatives of $w$ with respect to $x$.
	
	The successive practical steps of the method are the following.
	
	(I) Determine the possible values of $m$ and $n$ by balancing two or more terms of the PDE and expressing that they dominate the other terms.
	
	(II) Substitute the decompositions \eqref{gwy}-\eqref{gwt} into the PDE and integrability condition \eqref{ic}, then obtain the expressions for $w$ and its derivatives in the form of an overdetermined systems. 
	
	(III) Determine $F$ and $G$ by solving systems obtained in step II, whose solution yields the desired decomposition solution.

Examples are always a good starting point.
Consider the BKP equation
\begin{eqnarray}\label{bkp}
u_{xt} +(u_{x4} +15uu_{x2} +15u^3-15uv-5u_{xy})_{xx} -5u_{yy} =0, v_x =u_y, 
\end{eqnarray}
where $u_x = \partial_xu, u_{x2} = \partial_x^2u, u_{x3} = \partial_x^3u,\dots,$  which is a (2+1)-dimensional generalization of the Sawada-Kotera (SK) equation and exhibits many elegant integrable behavior \cite{ybib17,rbib21}. 
It is convenient to deal analytically with a potential function $w$, introduced by setting $u=w_x$, and it follows equation \eqref{bkp} that $w$ can be taken to satisfy the equation
\begin{eqnarray}\label{pbkp}
w_{xt} = 5w_{yy} - (w_{x5} + 15w_xw_{x3} + 15w_x^3 - 15w_xw_y - 5w_{xxy})_x. 
\end{eqnarray}

The CKP equation is defined as
\begin{eqnarray}\label{ckp}
&&9u_{xt}+(u_{x4}+15uu_{x2}+15u^3+\frac{45}{4}u_x^2-15uv-5u_{xy})_{xx}-5u_{yy}=0, v_x=u_y,
\\\label{pckp}
&&9w_{xt} ={5w_{yy}}  - ({w_{x5}} + 15w_xw_{x3} + {15w_x^3}+\frac{45w_{xx}^2}{4} - {15w_xw_y} -{5w_{xxy}})_x,
\end{eqnarray}
in which $u$ denotes the conservative field and $w$ the potential one.
When $u_y$ tends to zero, the reduced equation of the CKP equation \eqref{ckp} is the well-known Kaup–Kupershmidt (KK) equation \cite{ibib4}. The close connection between these two completely integrable fifth order equations is well documented \cite{jbib13}, nevertheless, despite their evident duality, equations \eqref{bkp} and \eqref{ckp} are fundamentally different. There is no scaling which reduces one to the other.

The application of the decomposition method to the potential BKP equation \eqref{pbkp} and potential CKP equation \eqref{pckp}  is described in sections II and III.
At the same time, the BTs and the Lax pairs for the potential BKP and potential CKP equations are found in a remarkably natural manner.  
Finally, some conclusions are given in part IV. Some lengthy and technical calculations have been moved to the Appendix.

\section{The decomposition solutions, B\"acklund transformation, Lax pair and linear superposition solutions of the potential BKP equation}

Analysis of the potential BKP equation \eqref{pbkp} shows  the leading orders $m=3,n=5$, in this case, $w_y$ and $w_t$ are expressible in terms of 
\begin{eqnarray}\label{wy}
w_y &= &F(x, y, t, w, w_x, w_{xx}, w_{xxx}), \\\label{wt}
w_t& = &G(x, y, t, w, w_x, w_{xx}, w_{xxx}, w_{xxxx}, w_{xxxxx}),
\end{eqnarray} 
and we therefore require
the equations \eqref{wy} and \eqref{wt} are subject to the integrability condition
\begin{eqnarray}\label{cc}
F_{t}=G_{y},
\end{eqnarray}
whenever equation \eqref{pbkp}  is satisfied.
The derivation of $F$ and $G$ is elementary but somewhat lengthy and can be found in the Appendix.  The result of calculation leads directly to the following theorems.

\textbf{Theorem 1.} If $w$ is a solution of the consistent variable coefficient potential Korteweg-de Vries (KdV) decomposition system 
\begin{eqnarray}\label{th1_1}
v_y &=&v_{xxx}-\frac{a}{2}v_x^2+\frac{m_0}{a}x+n_y,\\\nonumber
v_t&=&9v_{xxxxx}-15av_xv_ {xxx}-\frac{15a}{2}v_{xx}^2+\frac{5a^2}{2}v_x^3+[15m_2+15m_1y-(5m_0^2+\frac{1}{2}m_{0t})y^2-5m_0x]v_x\\\nonumber
&-&5m_0v+[(10m_0^2+m_{0t})y-15m_1]\frac{x}{a}+\frac{y^3}{6a}m_0(10m_0^2+m_{0t})-\frac{15y^2}{2a}m_0m_1-\frac{15y}{a}m_0m_2\\\label{th1_2}
&+&5m_0n+n_t+m_3,\\\label{th1_3}
w_y&=&w_{xxx}+3w_x^2+av_xw_x+\frac{a^2}{6}v_x^2-\frac{1}{3}m_0(x+m_0y^2)-\frac{m_{0t}y^2}{30}+m_1y+m_2,\\\nonumber
w_t&=&9w_{xxxxx}+15(av+6w)_xw_{xxx}+5a(av+3w)_xv_{xxx}+15(av+3w)_{xx}w_{xx}+\frac{15}{2}w_x[(av+3w)_x^2+3w_x^2]\\\nonumber
&+&\frac{5a^2}{2}v_{xx}^2+w_x(15m_2+15m_1y-5m_0^2y^2-5m_0x-\frac{1}{2}m_{0t}y^2)-5m_0w-\frac{xy}{3}(m_{0t}+10m_0^2)+5m_1x\\\label{th1_4}
&-&\frac{y^3}{90}(m_{0tt}+30m_0m_{0t}+100m_0^3)+\frac{y^2}{2}(m_{1t}+10m_0m_1)+(m_{2t}+10m_0m_2)_y+m_4,
\end{eqnarray}
where $m_i, i = 0, 1,\dots, 4$ are arbitrary functions of $t, n$ is an arbitrary function of $y, t$ and $a$ is an arbitrary constant, then $w$ is a solution of the potential BKP equation \eqref{pbkp}.

The arbitrary parameter $a$ determines that $v$ defined by equations \eqref{th1_1}-\eqref{th1_2} may be a decomposition solution of potential BKP equation \eqref{pbkp} as well.
The reader may care to show $v$  may indeed be transformed to a solution of potential BKP equation \eqref{pbkp} when $a=-3$ and $a=-6$.
Generalizing, we can state the following.

\textbf{Corollary 1.} Under the decomposition of Theorem 1, assume that the constant $a$ and the function $n$ be as
\begin{eqnarray}\label{co1}
a=-3, n=m_2y+ \frac{1}{2}m_1y^2 - \frac{1}{90}(10m_0^2 +m_{0t})y^3 +n_0(t),
\end{eqnarray}
then, both $v$ and $w$ given by \eqref{th1_1}-\eqref{th1_4} are solutions of the potential BKP equation \eqref{pbkp}.

We record another corollary of Theorem 1.

\textbf{Corollary 2.} The functions $v$ and $w$ fixed by \eqref{th1_1}-\eqref{th1_4} in the statement of Theorem 1 with the additional condition
\begin{eqnarray}\label{co2}
a=-6, n=m_2y+n_0(t), m_0 =m_1 =0
\end{eqnarray}
satisfy the potential BKP equation \eqref{pbkp} as well.

Note that the results in Corollaries 1 and 2 indicate that potential BKP equation \eqref{pbkp} possesses the variable coefficient KdV decomposition solutions \eqref{th1_1}-\eqref{th1_2} with conditions \eqref{co1} and/or \eqref{co2} \cite{xbib24}. 
On the other hand, the variable coefficient KdV decompositions \eqref{th1_3}-\eqref{th1_4} with \eqref{co1} and/or \eqref{co2} form BTs which connect two solutions $v$ and $w$ of the potential BKP equation \eqref{pbkp}.


A further consequence of decomposition is that we can establish a general BT associated with potential BKP \eqref{pbkp}. This result is formulated as the following theorem.

\textbf{Theorem 2. (B\"acklund Transformation)} The decomposition relation
\begin{eqnarray}\label{bt1}
p_y&=&p_{xxx}+\frac{3}{2}(w+v)_{xx}p+\frac{3}{2}(w_x^2-v_x^2)+\frac{1}{4}(p^3)_x, p\equiv w-v,\\\nonumber
p_t&=&9p_{xxxxx}+\frac{9}{16}(p^5)_x+\frac{45}{8}(w+v)_{xx}p^3-\frac{45}{8}(3v_x^2-3w_x^2-2p_{xxx})p^2+\frac{15}{4}(9v_x^2+9w_x^2-6v_xw_x\\\nonumber
&+&4v_{xxx}+4w_{xxx}+2v_y+2w_y)_xp-15w_{xxx}(v-4w)_x-15v_{xxx}(4v-w)_x+45w_{xx}^2-45v_{xx}^2\\\label{bt2}
&+&\frac{45}{2}p_x(v_x^2+w_x^2)+\frac{15}{2}(w+v)_yp_x,
\end{eqnarray}
constitutes a BT between two solutions $w$ and $v$ of the potential BKP equation \eqref{pbkp}.

It is easily verified by direct calculation that this system \eqref{bt1}-\eqref{bt2} is integrable and that $v$ satisfies equation \eqref{pbkp}. 
It is worth noting that the BT  has the form of a conservation law
\begin{eqnarray*}
p_y&=&[p_{xx}+\frac{3}{2}p(w_x+v_x)+\frac{p^3}{4}]_x,\\
p_t&=&[9p_{xxxx}+\frac{9p^5}{16}+\frac{45}{8}(w+v)_xp^3+\frac{45}{4}p^2p_{xx}+(15(w+v)_{xxx}+\frac{45}{2}(w_x^2+v_x^2)+\frac{15}{2}(w+v)_y)p\\&+&\frac{45}{2}(p_x(w+v)_x)_x]_x.
\end{eqnarray*}
 
 Finding the BT of a PDE is equivalent to finding its Lax pair \cite{mbib16,mbib17}.
Therefore, further consideration of the BT for the construction of the Lax pair seems warranted.
With appeal to Theorem 2, we can obtain the well-known Lax pair  \cite{bbib2}
\begin{eqnarray}
\psi_y&=&\psi_{xxx}+3w_x\psi_x,\label{Lx}\\
\psi_t&=&9\psi_{xxxxx}+45 w_x \psi_{xxx}+45 w_{xx}\psi_{xx} +15(2 w_{xxx}-3w_x^2+w_y)\psi_x\label{Lt}
\end{eqnarray}
of the potential BKP equation \eqref{pbkp}  by  substitution of the transformation
$v=w+\ln(\psi^2)_{x}$ into the BT \eqref{bt1}-\eqref{bt2} 
as may be verified by direct calculation.
The elimination of $w$ is made between Lax pair, and results in 
the Schwarzian form of the potential BKP equation \eqref{pbkp}
\begin{eqnarray}
S_{xxx}+C_x+4S^2_x-5K^2_x+5(S_xK-S_y-K_y)=0
\end{eqnarray}
with Schwarzian derivatives $S=\frac{\psi_{xxx}}{\psi_x}-\frac{3\psi_{xx}^2}{2\psi_x^2}, C=\frac{\psi_t}{\psi_x}$ and $K=\frac{\psi_y}{\psi_x}$.

Notice that 1. Each of two Corollaries and Theorem 2 takes the form of a system of coupled differential equations for two unknown functions, and all imply that if one of the two functions is a solution of the potential BKP \eqref{pbkp}, then also the second function solves the same equation. 2. The BT \eqref{bt1}-\eqref{bt2} implies a type of the Sharma–Tasso–Olver (STO) decomposition solution of the potential BKP equation \eqref{pbkp} when $v=0$ \cite{xbib24}, also presents a means of finding the Lax pair, then Schwarzian equation.

\textbf{Theorem 3.} Let the function $w$ be a solution to the consistent variable coefficient Svinolupov Sokolov (SS) system
\begin{eqnarray}
w_y&=&w_{xxx}+\frac{3w_{xx}^2}{2W}+\frac{3}{2}w_x^2-\frac{27}{4}M^2+\frac{3}{2}m_1x+\frac{3y}{20}(m_1y+2m_2)_t+m_3,\\\nonumber
w_t&=&9w_{xxxxx}+45w_xw_{xxx}+\frac{45}{2}w_x^3+\frac{3}{4}[30m_1x-135M^2+3y(m_1y+2m_2)_t+20m_3]w_x+\frac{45}{2}m_1w\\\nonumber
&-&\frac{3}{2}(45m_1M-m_{1t}y-m_{2t})x+\frac{y^2}{20}(m_1y+3m_2)_{tt}-\frac{27}{4}y(m_1y+2m_2)_tM+y(m_{3t}-45m_1m_3)+m_4\\
&+&\frac{45}{2W}[2w_{xx}w_{xxxx}+w_{xxx}^2+(3M+w_x)w_{xx}^2]+\frac{180}{W^2}w_{xx}^2w_{xxx}+\frac{315}{2W^3}w_{xx}^4+\frac{405}{4}M^3,
\end{eqnarray}
where $M \equiv m_1y+m_2, W \equiv 3M-2w_x$ and $m_i, i = 1, 2, 3, 4$ are arbitrary functions of $t$, then $w$ is a solution of the potential BKP equation \eqref{pbkp}.

\textbf{Theorem 4.} The potential BKP equation \eqref{pbkp} possesses the following variable coefficient KdV decomposition solution
\begin{eqnarray}
w_y&=&-\frac{1}{2}w_{xxx}-\frac{3}{2}w_x^2+6Mw_x+\frac{y}{10}(m_1y+2m_2)_t+m_1x-6M^2+m_3, M\equiv m_1y+m_2,\\\nonumber
w_t&=&-\frac{9}{4}w_{xxxxx}-\frac{45}{4}(2w_xw_{xxx}+w_{xx}^2+2w_x^3)+15m_1w+\frac{3}{2}y(m_1y+2m_2)_tw_x+15(6M^2+m_1x+m_3)w_x\\
&+&\frac{y^2}{30}(m_1y+3m_2)_{tt}-3(y^2m_{1t}+10m_1x+2ym_{2t})M+M_tx+(m_{3t}-30m_1m_3)y-60M^3+m_4,
\end{eqnarray}
where $m_i, i = 1, 2, 3, 4$ are arbitrary functions of $t$.

\textbf{Theorem 5.} The variable coefficient potential SK decomposition solution of the potential BKP equation \eqref{pbkp} possesses the form
\begin{eqnarray}\label{bysk}
w_y&=&Mw_x+\frac{y}{30}(M+m_2)_t+\frac{1}{3}m_1x-\frac{1}{2}M^2+m_3, M\equiv m_1y+m_2,\\\nonumber
w_t&=&-w_{xxxxx}+5(M-3w_x)w_{xxx}+15(M-w_x)w_x^2+(\frac{1}{2}y^2m_{1t}+ym_{2t}+5m_1x-\frac{5}{2}M^2+15m_3)w_x\\\nonumber
&+&5m_1w+\frac{y^2}{90}(m_1y+3m_2)_{tt}-\frac{y}{3}(My-x)m_{1t}+\frac{1}{3}(x-2M_y)m_{2t}-\frac{10}{3}m_1(Mx+3m_3y)\\\label{btsk}&+&m_{3t}y+m_4
\end{eqnarray}
with $m_i, i = 1, 2, \dots, 4$ being arbitrary functions of $t$.

Suppose that arbitrary functions $m_i=0, i = 1, 2, \dots, 4$, then by Theorems 3 to 5, three special decompositions correspond to constant coefficient SS , KdV and SK decomposition solutions of potential BKP equation \eqref{pbkp}, respectively \cite{xbib24}.

If a BT represents one of the different aspects of the property of integrability for a PDE, the existence of a linear combination principle allows further to build explicitly some classes of solutions depending on decomposition system \eqref{th1_1}-\eqref{th1_2}. 
The linear combination solution of the potential BKP equation \eqref{pbkp}  from the result of the decomposition is established.

\textbf{Theorem 6.} If $v_1$ and $v_2$ are solutions of the variable coefficient potential KdV decompositions
\begin{eqnarray}
v_{1y}&=&v_{1xxx}-\frac{1}{2}a_1v_{1x}^2+\frac{1}{a_1}m_0x+n_y, M\equiv m_1y+m_2,\\\nonumber
v_{1t}&=&9v_{1xxxxx}+\frac{5}{2}a_1^2v_{1x}^3-\frac{15}{2}a_1(2v_{1x}v_{1xxx}+v_{1xx}^2)+[15M-5m_0x-\frac{1}{2}(10m_0^2+m_{0t})y^2])v_{1x}\\\nonumber
&-&5m_0v_1+5m_0n+n_t+m_3+\frac{1}{a_1}\{[(10m_0^2+m_{0t})y-15m_1]x+\frac{5}{3}y^3m_0^3\\
&+&\frac{1}{6}m_0m_{0t}y^3-\frac{15}{2}m_0(m_1y+2m_2)y\},
\end{eqnarray}
and 
\begin{eqnarray}
v_{2y}&=&v_{2xxx}-\frac{1}{2}a_2v_{2x}^2+\frac{1}{a_2}(\frac{1}{5}m_{0t}y^2-a_1n_y+m_0x+2y^2m_0^2-6M),\\\nonumber
v_{2t}&=&9v_{2xxxxx}+\frac{5}{2}a_2^2v_{2x}^3-\frac{15}{2}a_2(2v_{2x}v_{2xxx}+v_{2xx}^2)-\frac{1}{2}y^2v_{2x}m_{0t}+(15M-5m_0^2y^2-5m_0x)v_{2x}-5m_0v_2\\\nonumber
&+&m_4+\frac{1}{a_2}[\frac{1}{15}m_{0tt}y^3+\frac{1}{6}(6xy+11m_0y^3)m_{0t}-3m_{1t}y^2-6m_{2t}y-a_1n_t+5(2m_0^2y-3m_1)x\\
&+&5y^3m_0^3-\frac{45}{2}m_0m_1y^2-45m_0m_2y-5m_0a_1n],
\end{eqnarray}
respectively, then their linear combination
\begin{eqnarray}\label{sum}
w=-\frac{1}{6}(a_1v_1+a_2v_2)
\end{eqnarray}
is a solution of the potential BKP equation \eqref{pbkp}.

Generally, the functions $v_1$ and $v_2$ in Theorem 6 are not the solutions of the potential BKP equation \eqref{pbkp} except in certain special cases mentioned in Corollaries 1 and 2, whereas the linear combination \eqref{sum} of $v_1$ and $v_2$ solves the potential BKP equation \eqref{pbkp}.
The linear superposition solutions with fixed $a_1$ and $a_2$ in \cite{xbib24} are particular cases of Theorem 6.

\section{The Decomposition solutions, B\"acklund transformation and Lax pair of the potential CKP equation}

As a second illustration, we repeat the entire procedure in the previous section to find the decomposition solutions and B\"acklund transformation of the potential CKP equation \eqref{pckp}.
Similarly, a straightforward calculation determines the following theorems.

\textbf{Theorem 7. (B\"acklund Transformation)} Let $v$ be any solution of the potential CKP equation \eqref{pckp}, a different solution $w$ of equation \eqref{pckp} is then defined by the B\"acklund transformation 
\begin{eqnarray}\label{cbt1}
	p_y&=&[p_{xx}-\frac{3p_x^2}{4p}+\frac{3}{2}p(w_x+v_x)+\frac{p^3}{4}]_x, \ p=w-v,\\\nonumber
	p_t&=&[p_{xxxx}-\frac{5}{2p}p_xp_{xxx}+\frac{5p}{3}(w+v)_{xxx}-\frac{5}{4p}p_{xx}^2+(\frac{5p^2}{4}+\frac{5p_x^2}{p^2}+\frac{5}{2}(w+v)_x)p_{xx}+\frac{5}{4}p_x(w+v)_{xx}\\\label{cbt2}&-&\frac{35p_x^4}{16p^3}-\frac{15(w+v)_x}{8p}p_x^2+\frac{15}{8}p(w_x^2+v_x^2)+\frac{p^5}{16}+\frac{5p}{6}(w+v)_y+\frac{5p^3}{8}(w+v)_x+\frac{5}{4}pw_xv_x]_x.
\end{eqnarray}

This BT \eqref{cbt1} and \eqref{cbt2} appears to be a new result and is again in a conservation form. 
The Lax pair arises from the BT, which results in two equations  
\begin{eqnarray}\label{clax1}
\psi_y&=&\psi_{xxx}+3w_x\psi_x+\frac32w_{xx}\psi,\\\label{clax2}
\psi_t&=&\psi_{xxxxx}+5w_x\psi_{xxx}+\frac{15}{2}w_{xx}\psi_{xx}+5(\frac{7}{6}w_{xxx}+w_x^2+\frac{w_y}{3})\psi_x+5(\frac{w_{xxxx}}{3}+w_xw_{xx}+\frac{w_{xy}}{6})\psi
\end{eqnarray}
 followed by the relation
	$v=w+\ln(\partial^{-1}_x \psi^2)_x$.

When $v=0$, the BT is reduced to the modified STO decomposition solution 
\begin{eqnarray}
w_y&=&[w_{xx}-\frac{3w_x^2}{4w}+\frac{3}{2}ww_x+\frac{w^3}{4}]_x,\\\nonumber
w_t&=&[w_{xxxx}-\frac{5}{2w}w_xw_{xxx}+\frac{5}{3}ww_{xxx}-\frac{5}{4w}w_{xx}^2+5(\frac{w^2}{4}+\frac{w_x^2}{w^2}+\frac{3}{4}w_x)w_{xx}-\frac{35w_x^4}{16w^3}-\frac{15w_x^3}{8w}\\&+&\frac{15ww_x^2}{8}+\frac{w^5}{16}+\frac{5ww_y}{6}+\frac{5w^3w_x}{8}]_x
\end{eqnarray}
of the potential CKP equation \eqref{pckp}. This very special  modified STO decomposition is linked to the STO decomposition
\begin{eqnarray}\label{csto1}
f_y&=&[f_{xx}+\frac{1}{4}f^3+\frac{3}{2}(f_{x}f)]_x,\\\label{csto2}
f_t&=&[f_{xxxx}+5f_xf_{xx}+5(\frac{ff_{xxx}}{2}+\frac{f^2f_{xx}}{2}+\frac{f^3f_x}{4}+\frac{3ff_x^2}{4})+\frac{f^5}{16}]_x
\end{eqnarray}
by transformation $w_x=fw-w^2$, which is exactly the same as the STO decomposition of the potential BKP equation \eqref{pbkp}.

\textbf{Theorem 8. } The function $w$ is a solution of the potential CKP equation \eqref{pckp} provided that $w$ satisfies the consistent variable coefficient SS system
\begin{eqnarray}
w_y&=&w_{xxx}+\frac{3w_{xx}^2}{2W}+{3}w_x^2-\frac{9Mw_x}{2}-\frac{27}{8}M^2+\frac{3}{2}m_1x+\frac{27y}{20}(m_1y+2m_2)_t+\frac{27m_2^2}{8}+m_3,\\\nonumber
w_t&=&w_{xxxxx}+\frac{5w_{xx}w_{xxxx}}{W}+\frac{5w_{xxx}^2}{2W}+(\frac{20w_{xx}^2}{W^2}+10w_x-\frac{15M}{2})w_{xxx}+\frac{35w_{xx}^4}{2W^3}+\frac{5(3M+4w_x)w_{xx}^2}{4W}\\\nonumber
&+&10w_x^3-\frac{45M}{2}w_x^2+(\frac{45}{8}M^2+\frac{45m_2^2}{8}+\frac{5m_1x}{2}+\frac{5m_3}{3})w_x+\frac{(9w_xy^2+6xy-27y^2M)m_{1t}}{4}\\\nonumber
&+&\frac{(18w_xy+6x-27y(2m_1y+m2))m_{2t}}{4}+\frac{9y^2(m_1y+3m_2)_{tt}}{20}-\frac{15m_1Mx}{2}+\frac{45M^3}{4}-\frac{135m_1m_2^2y}{8}\\&+&\frac{5m_1w}{2}-5m_1m_3y+m_{3t}y+m_4,
\end{eqnarray}
where $M \equiv m_1y+m_2, W \equiv 3M-2w_x$ and $m_i, i = 1, 2, 3, 4$ are arbitrary functions of $t$ .

\textbf{Theorem 9.} The potential CKP equation \eqref{pckp} possesses the following variable coefficient KdV decomposition solution
\begin{eqnarray}
w_y&=&\frac{1}{4}w_{xxx}+\frac{3}{2}w_x^2+\frac{3m_0x}{5}+\frac{27y^2(m_{0t}-2m_0^2)}{50}+M, M\equiv m_1y+m_2,\\\nonumber
w_t&=&\frac{1}{16}w_{xxxxx}+\frac{5}{8}(2w_xw_{xxx}+w_{xx}^2+4w_x^3)+(m_0x-\frac{9m_0^2y^2}{5}+\frac{5}{3}M+\frac{9m_{0t}y^2}{10})w_x+m_0w\\\nonumber
&+&((\frac{3m_{0t}}{5}-\frac{6m_0^2}{5})y+\frac{5m_1}{9})x+(\frac{18m_0^3}{25}-\frac{27m_0m_{0t}}{25}+\frac{9m_{0tt}}{50})y^3+(\frac{m_{1t}}{2}-m_0m_1)y^2\\&+&(m_{2t}-2m_0m_2)y+m_3,
\end{eqnarray}
where $m_i, i = 1, 2, 3, 4$ are arbitrary functions of $t$.

\textbf{Theorem 10.} The variable coefficient potential KK decomposition solution of the potential CKP equation \eqref{pckp} possesses the form
\begin{eqnarray}
w_y&=&Mw_x+\frac{3y}{10}(M+m_2)_t+\frac{1}{3}m_1x-\frac{1}{2}M^2+\frac{m_2^2}{2}+m_3, M\equiv m_1y+m_2,\\\nonumber
w_t&=&-\frac{w_{xxxxx}}{9}+\frac{5}{9}(M-3w_x)w_{xxx}-\frac{5w_{xx}^2}{4}-\frac{5w_x^3}{3}+\frac{5M}{3}w_x^2+(\frac{5m_1x}{9}-\frac{5M^2}{18}+\frac{5m_2^2}{6}+\frac{5m_3}{3}\\\nonumber&+&\frac{m_{1t}y^2}{2}+m_{2t}y)w_x+\frac{5m_1w}{9}+\frac{y^2}{10}(m_1y+3m_2)_{tt}-\frac{m_{1t}y^2M}{3}+\frac{xM_t}{3}+\frac{m_{2t}y(m_2-2m_1y)}{3}\\&+&m_{3t}y-\frac{10m_1xM}{27}-\frac{5m_1y(m_2^2+2m_3)}{9}+m_4
\end{eqnarray}
with $m_i, i = 1, 2, \dots, 4$ being arbitrary functions of $t$.

The proof of Theorems 7-10 is of the same form as that presented in the Appendix for the potential BKP equation \eqref{pbkp}. The calculation is omitted here.

 We have attempted but failed to provide the linear superposition solution of the potential CKP equation \eqref{pckp} from the decomposition solutions.

\section{Conclusions}

In this study, which extends the work of a previous paper \cite{xbib24}, we have presented a modified formal variable separation approach for deriving decomposition solutions of the potential BKP and CKP equations. 
This method is heuristic, it gives decomposition solutions of the (2+1)-dimensional PDEs in a unified way. The result makes clear that we establish the connection between the potential BKP and CKP equations with several classic integrable systems.
One might therefore hope to deduce the solutions of the potential BKP and CKP equations from solutions of the classical integrable systems.

Specifically, we are able to construct the BTs from decomposition.
It is widely accepted that the existence of a BT serves as a sign to the integrability of a PDE. While the BTs for the (1+1)-dimensional SK and KK equations have been known for some time \cite{mbib16}, the BTs for the (2+1)-dimensional SK \eqref{bkp} and KK \eqref{ckp} equations  have not previously been reported. The BTs of the potential BKP and CKP equations are given here in explicit forms for the first time. 
Of course, other standard analytic techniques for obtaining BTs of nonlinear PDEs are available, but, the construction of the BTs from decomposition is a remarkably straightforward way.

In the case of the potential BKP and CKP equations, the system defining the BTs found has two properties. First, it is linearizable since it results in the Lax pair. Second, it has a conservation form. 
Observe that the solution to the potential BKP equation in Theorem 6 also has a unique property: the well-known linear superposition principle holds if the parameters are taken on some special values.

It is note that yet despite the close resemblance of the potential BKP and CKP equations, they are fundamentally different. There is no scaling which transformations one equation into the other. 
Whereas in view of the system \eqref{csto1}-\eqref{csto2}, it is seen that the STO decomposition links the potential BKP and CKP equations.

As an elementary tool for constructing solutions of PDEs, decomposition is as straightforward to apply as us believe. We hope to expand on some of these theoretical aspects in future work.




\section*{Acknowledgement}
The work was sponsored by the National Natural Science Foundations of China (Nos. 11975131, 11435005), K. C. Wong Magna Fund in Ningbo University, the Natural Science Foundation of Zhejiang Province No. LQ20A010009. 

\bibliographystyle{elsarticle-num}
\bibliography{ref}

\begin{thebibliography}{10}
\expandafter\ifx\csname url\endcsname\relax
  \def\url#1{\texttt{#1}}\fi
\expandafter\ifx\csname urlprefix\endcsname\relax\def\urlprefix{URL }\fi
\expandafter\ifx\csname href\endcsname\relax
  \def\href#1#2{#2} \def\path#1{#1}\fi

\bibitem{abib14}
A.~S. Fokas, M.~J. Ablowitz, {On the inverse scattering transform of
  multidimensional nonlinear equations related to first order systems in the
  plane}, Journal of Mathematical Physics 25 (1984) 2494--2505.

\bibitem{rbib22}
R.~Hirota, {A new form of B\"acklund transformations and its relation to the
  inverse scattering problem}, Prog. theor. Phys. 52 (1974) 1498--1512.

\bibitem{vbib4}
V.~O. Vakhnenko, E.~J. Parkes, A.~J. Morrison, {A B\"acklund transformation and
  the inverse scattering transform method for the generalised Vakhnenko
  equation}, Chaos, Solitons and Fractals 17 (2003) 683--692.

\bibitem{abib23}
A.~S. Fokas, R.~L. Anderson, {Group theoretical nature of B\"acklund
  transformations}, Lett. Math. Phys. 3 (1979) 117--126.

\bibitem{jbib3}
J.~Weiss, M.~Tabor, G.~Carnevale, {The Painlev\'e property for partial
  differential equations}, Journal of Mathematical Physics 24 (1983) 522--526.

\bibitem{jbib9}
J.~Weiss, {The {P}ainlev\'{e} property for partial differential equations.
  {II}: {B}\"{a}cklund transformation, {L}ax pair, and the {S}chwarzian
  derivative}, Journal of Mathematical Physics 24 (1983) 1405--1413.

\bibitem{rbib3}
R.~Conte, M.~Musette, {Painlev\'{e} analysis and {B}\"{a}cklund transformation
  in the {K}uramoto-{S}ivashinsky equation}, Journal of Physics A: Mathematical
  and General 22 (1989) 169--177.

\bibitem{sbib8}
S.~Y. Lou, {Painlev\'e test for the integrable dispersive long wave equations
  in two space dimensions}, Physics Letters A 176 (1993) 96--100.

\bibitem{xbib1}
X.~N. Gao, S.~Y. Lou, X.~Y. Tang, {Bosonization, singularity analysis, nonlocal
  symmetry reductions and exact solutions of supersymmetric KdV equation},
  Journal of High Energy Physics 05 (2013) 29.

\bibitem{ebib14}
E.~G. Fan, {Two new applications of the homogeneous balance method}, Phys.
  Lett. A 265 (2000) 353--357.

\bibitem{ybib4}
Y.~Jin, M.~Jia, S.~Y. Lou, {B\"{a}cklund transformations and interaction
  solutions of the {B}urgers equation}, Chinese Physics Letters 30 (2013)
  020203.

\bibitem{wbib25}
W.~X. Ma, A.~Abdeljabbar, {A bilinear B\"acklund transformation of a
  (3+1)-dimensional generalized KP equation}, Appl. Math. Lett. 25 (2012)
  1500--1504.

\bibitem{gbib16}
G.~L. Lamb, {B\"acklund transformations for certain nonlinear evolution
  equations}, J. Math. Phys. 15 (1974) 2157--2165.

\bibitem{mbib2}
M.~C. Nucci, {Pseudopotentials, {L}ax equations and {B}\"{a}cklund
  transformations for nonlinear evolution equations}, Journal of Physics A:
  Mathematical and General 21 (1988) 73--79.

\bibitem{mbib15}
M.~Wadati, H.~Sanuki, K.~Konno, {Relationships among inverse method, B\"acklund
  transformation and an infinite number of conservation laws}, Prog. Theor.
  Phys. 53 (1975) 419--436.

\bibitem{jbib10}
J.~Satsuma, D.~J. Kaup, {A {B}\"{a}cklund transformation for a higher order
  Korteweg-de Vries equation}, Journal of the Physical Society of Japan 43
  (1977) 692--697.

\bibitem{sbib42}
S.~Y. Lou, L.~L. Chen, {Formal variable separation approach for nonintegrable
  models}, J. Math. Phys. 40 (1999) 6491--6500.

\bibitem{sbib43}
S.~Y. Lou, X.~Y. Tang, J.~Lin, {Exact solutions of the coupled KdV system via a
  formally variable separation approach}, Commun. Theor. Phys. 36 (2001)
  145--148.

\bibitem{xbib26}
X.~Y. Tang, S.~Y. Lou, {A variable separation approach to solve the integrable
  and nonintegrable models: coherent structures of the (2+1)-dimensional KdV
  equation}, Commun. Theor. Phys. 38 (2002) 1--8.

\bibitem{ybib17}
Y.~Li, R.~X. Yao, Y.~R. Xia, S.~Y. Lou, {Plenty of novel interaction structures
  of soliton molecules and asymmetric solitons to (2+1)-dimensional
  Sawada-Kotera equation}, Commun. Nonlinear Sci. Numer. Simulat. 100 (2021)
  105843.

\bibitem{rbib21}
R.~X. Yao, Y.~Li, S.~Y. Lou, {A new set and new relations of multiple soliton
  solutions of (2+1)-dimensional Sawada-Kotera equation}, Commun. Nonlinear
  Sci. Numer. Simulat. 99 (2021) 105820.

\bibitem{ibib4}
I.~Loris, {On reduced CKP equations}, Inverse Problems 15 (1999) 1099--1109.

\bibitem{jbib13}
J.~Weiss, {On class of integrable systems and the Painlev\'e property}, Journal
  of Mathematical Physics 25 (1984) 13--24.

\bibitem{xbib24}
X.~Z. Hao, S.~Y. Lou, {Decompositions and linear superpositions of B-type
  Kadomtsev-Petviashvili equations}, Math. Meth. Appl. Sci. (2022) 5774--5796.

\bibitem{mbib16}
M.~Musette, R.~Conte, {B\"acklund transformation of partial differential
  equations from the Painlev\'e-Gambier classification. I. Kaup-Kupershmidt
  equation}, J. Math. Phys. 39 (1998) 5617--5630.

\bibitem{mbib17}
M.~Musette, C.~Verhoeven, {Nonlinear superposition formula for the
  Kaup-Kupershmidt partial differential equation}, Phys. D 144 (2000) 211--220.

\bibitem{bbib2}
B.~G. Konopelchenko, V.~G. Dubrovsky, {Some new integrable nonlinear evolution
  equations in 2+1 dimensions}, Phys. Lett. A 102 (1984) 15--17.

\end{thebibliography}
\section*{Appendix }

We illustrate the decomposition procedure for the case of the potential BKP equation \eqref{pbkp}.
\begin{proof}
		Substituting \eqref{wy} and \eqref{wt} with $m>3$ into the potential BKP equation \eqref{pbkp}, plus the decomposition compatibility condition \eqref{cc}, one can find that 
		the decomposition does not hold generally for $m>3$.
		Therefore, we take $m=3$ and then $n=5$ in the decomposition relations \eqref{wy} and \eqref{wt}. Substituting \eqref{wy} and \eqref{wt} into \eqref{pbkp} gives 
	\begin{eqnarray}
	w_{x6}(1-5F_{x_3}^2 -5F_{x_3}+G_{x_5})+W=0, \label{x6}
	\end{eqnarray}
	where $F=F(x,y,t,w, w_x,  w_{x2}, w_{x3})\equiv F(x,y,t,x_0, x_1, x_2, x_3)$, $G=G(x,y,t,w, w_x, w_{x2},w_{x3},w_{x4}, w_{x5})\equiv G(x,y,t,x_0, x_1, x_2, x_3, x_4, x_5)$, and $W=W(x,y,t,x_0, x_1, \ldots, x_5)$ is a complicated expression of $x,y,t,x_0, x_1, \ldots, x_5$. Vanishing coefficient of $w_{x6}$, we have
	\begin{eqnarray}\label{GG1}
	G=(5F_{x_3}+5F_{x_3}^2-1)x_5+G_1,
	\end{eqnarray}
	where $G_1=G_1(x,y,t,x_0, x_1, \ldots, x_4)$ is a function of $\{x,y,t,x_0, x_1, \ldots, x_4\}$.
	By using the relation \eqref{GG1}, \eqref{x6} is changed to
	\begin{eqnarray}\nonumber
	&&w_{x5}[G_{1x_4}-5(F_{x_3}+2)(x_1F_{x_0x_3}+x_2F_{x_1x_3}+x_3F_{x_2x_3}+x_4F_{x_3x_3})-5F_{x_2}(1+2F_{x_3})-5F_{xx_3}(2+F_{x_3})]\\\label{x5}
	&&+W_1=0,
	\end{eqnarray}
	where $W_1=W_1(x,y,t,x_0, x_1, \ldots, x_4)$ is $w_{x5}$ independent. Eliminating the coefficient of $w_{x5}$ yields
	\begin{eqnarray}\nonumber
	G_{1}&=&5 (F_{x_3}+2)(\frac12 x_4F_{x_3x_3}+x_3 F_{x_2x_3}+x_2F_{x_1x_3}+x_1F_{x_0x_3})x_4 +5x_4F_{x_2}(1+2F_{x_3})\\
	&+&5x_4F_{xx_3}(2+F_{x_3})+G_2, \label{GG2}
	\end{eqnarray}
	with $G_2\equiv G_2(x,y,t,x_0, x_1, x_2, x_3)$ being a function of $\{x,y,t,x_0, x_1, x_2, x_3\}$.
	
	Similarly, substituting the decomposition \eqref{wy} and \eqref{wt} with $m=3,\ n=5$, the results \eqref{GG1} and \eqref{GG2} into the consistent condition \eqref{cc}, we have
	\begin{eqnarray}
	5w_{x7}(1-F_{x_3})^2(x_1F_{x_0x_3} +x_2F_{x_1x_3} +x_3F_{x_2x_3} +{x_4}F_{x_3x_3}+F_{xx_3})+\Gamma=0, \label{x7}
	\end{eqnarray}
	where $\Gamma=\Gamma(x,y,t,x_0, \ldots, x_6)$ is a $w_{x7}$ independent function of the lower-order differentiations of $w$ with respect to $x$.
	Vanishing the coefficient of $w_{x7}w_{x4}$ in \eqref{x7}, we get
	\begin{eqnarray}
	F=F_1(x,y,t,x_0, x_1, x_2)x_3+H(x,y,t,x_0, x_1, x_2). \label{rF}
	\end{eqnarray}
	Substituting \eqref{rF} into \eqref{x7} and requiring the coefficient of $w_{x7}$ being zero result $F_1(x_0,\ x_1,\ x_2)=H_1(y,t)$. Thus, we have
	\begin{eqnarray}
	F=H_1(y,t) x_3+H(x,y,t,x_0, x_1, x_2). \label{rF0}
	\end{eqnarray}
	Taking account of \eqref{rF0}, \eqref{x5} becomes
	\begin{eqnarray}
	&&w_{x4}\big[G_{2x_3}-5(H_1+2)x_3 H_{x_2x_2}-5(H_1+2)x_2 H_{x_1x_2}+5x_1(3-H_1 H_{x_0x_2} -3H_1-2H_{x_0x_2})\nonumber\\
	&&-10H_1 H_{x_1}-5 H_{x_2}^2-5 H_{x_1}-5H_1H_{xx_2}-10H_{xx_2}\big]+W_2=0 \label{x4}
	\end{eqnarray}
	with $W_2=W_2(x,y,t,x_0, x_1, x_2,x_3)$. Vanishing the coefficient of $w_{x4}$ in \eqref{x4} leads to
	\begin{eqnarray}
	G_{2}&=&5\big[\frac{H_1+2}2x_3 H_{x_2x_2}+(H_1+2)x_2 H_{x_1x_2}-x_1(3-H_1 H_{x_0x_2} -3H_1-2H_{x_0x_2})+2H_1 H_{x_1}\nonumber\\
	&&+ H_{x_2}^2+ H_{x_1}+H_1H_{xx_2}+2H_{xx_2}\big]x_3+J, \label{G2}
	\end{eqnarray}
	where $J=J(x,y,t,x_0, x_1, x_2)$. Up to now, the decomposition relation is simplified to
	\begin{eqnarray}
	w_y&&=H_1w_{x3}+H,\label{wy1}\\\nonumber
	w_t&&=(5H_1+5H_1^2-1)w_{x_5}+5H_{x_2}(1+2H_1)w_{x_4}+5\big[\frac{H_1+2}2w_{x_3} H_{x_2x_2}+(H_1+2)w_{x_2} H_{x_1x_2}\\
	&&-w_{x_1}(3-H_1 H_{x_0x_2} -3H_1-2H_{x_0x_2})+2H_1 H_{x_1}+ H_{x_2}^2+ H_{x_1}+H_1H_{xx_2}+2H_{xx_2}\big]w_{x_3}+J \label{wt1}
	\end{eqnarray}
	with  three undetermined functions $H_1=H_1(y,t), H=H(x,y,t,x_0,x_1, x_2)$ and $J=J(x,y,t,x_0, x_1, x_2)$.

	Inserting \eqref{wy1} and \eqref{wt1} into the potential BKP equation \eqref{pbkp} and the consistent condition \eqref{cc}, then, vanishing the coefficients of $w_{xk}$ for $k\geq 3$ leaves the  set of determining equations on $\{H, H_1, J\}$.	
	Solving this set of equations, for nontrivial solutions we shall have the following several cases for the unknowns.
	
	\bf \it Case 1. \rm When $H_1=0$, further calculation then leads to the expression
	\begin{align}\left\{
	\begin{aligned}
	\label{HJ_SK}
	H&=Mw_x+\frac{y}{30}(M+m_2)_t+\frac{1}{3}m_1x-\frac{1}{2}M^2+m_3, M\equiv m_1y+m_2,\\
J&=15(M-w_x)w_x^2+(\frac{1}{2}y^2m_{1t}+ym_{2t}+5m_1x-\frac{5}{2}M^2+15m_3)w_x+5m_1w+\frac{y^2}{90}(m_1y+3m_2)_{tt}\\&-\frac{y}{3}(My-x)m_{1t}+\frac{1}{3}(x-2M_y)m_{2t}-\frac{10}{3}m_1(Mx+3m_3y)+m_{3t}y+m_4
	\end{aligned}\right.
	\end{align}
	with $m_i, i = 1, 2, \dots, 4$ being arbitrary functions of $t$.

	\bf \it Case 2. \rm Taking $H_1=1$ in the result of this case, then one can find three different solutions. 
	
	The first one is  
	\begin{align}\left\{
	\begin{aligned}\label{HJ_KdV1}
	H&=\frac{3w_{xx}^2}{2W}+\frac{3}{2}w_x^2-\frac{27}{4}M^2+\frac{3}{2}m_1x+\frac{3y}{20}(m_1y+2m_2)_t+m_3,\\
	J&=\frac{45}{2}w_x^3+\frac{3}{4}[30m_1x-135M^2+3y(m_1y+2m_2)_t+20m_3]w_x+\frac{45}{2}m_1w-\frac{3}{2}(45m_1M-m_{1t}y\\&-m_{2t})x+\frac{y^2}{20}(m_1y+3m_2)_{tt}-\frac{27}{4}y(m_1y+2m_2)_tM+y(m_{3t}-45m_1m_3)+m_4\\&+\frac{45}{2W}[(3M+w_x)w_{xx}^2]+\frac{315}{2W^3}w_{xx}^4+\frac{405}{4}M^3,
	\end{aligned}\right.
	\end{align}
where $M \equiv m_1y+m_2, W \equiv 3M-2w_x$ and $m_i, i = 1, 2, 3, 4$ are arbitrary functions of $t$.

	The second one is
	\begin{equation}\left\{
	\begin{aligned}
	H& = 3w_x^2+av_xw_x+\frac{a^2}{6}v_x^2-\frac{1}{3}m_0(x+m_0y^2)-\frac{m_{0t}y^2}{30}+m_1y+m_2,\\
	 J &=5a(av+3w)_xv_{xxx}+15(av+3w)_{xx}w_{xx}+\frac{15}{2}w_x[(av+3w)_x^2+3w_x^2]+\frac{5a^2}{2}v_{xx}^2+w_x(15m_2\\&+15m_1y-5m_0^2y^2-5m_0x-\frac{1}{2}m_{0t}y^2)-5m_0w-\frac{xy}{3}(m_{0t}+10m_0^2)+5m_1x-\frac{y^3}{90}(m_{0tt}\\&+30m_0m_{0t}+100m_0^3)+\frac{y^2}{2}(m_{1t}+10m_0m_1)+(m_{2t}+10m_0m_2)_y+m_4
	\end{aligned}\right.
	\end{equation}
	with $v$ satisfying
	\begin{equation}\left\{
\begin{aligned}
v_y =&v_{xxx}-\frac{a}{2}v_x^2+\frac{m_0}{a}x+n_y,\\
v_t=&9v_{xxxxx}-15av_xv_ {xxx}-\frac{15a}{2}v_{xx}^2+\frac{5a^2}{2}v_x^3+[15m_2+15m_1y-(5m_0^2+\frac{1}{2}m_{0t})y^2-5m_0x]v_x\\
&-5m_0v+[(10m_0^2+m_{0t})y-15m_1]\frac{x}{a}+\frac{y^3}{6a}m_0(10m_0^2+m_{0t})-\frac{15y^2}{2a}m_0m_1-\frac{15y}{a}m_0m_2\\
&+5m_0n+n_t+m_3.
	\end{aligned}\right.
\end{equation}	

The third one is
	\begin{equation}\left\{
\begin{aligned}
	H&=-v_{xxx}+v_y+\frac{3}{2}(w+v)_{xx}(w-v)+\frac{3}{2}(w_x^2-v_x^2)+\frac{1}{4}[(w-v)^3]_x, p\equiv w-v,\\
	J&=v_t-9v_{xxxxx}+\frac{15}{2}pv_{xxxx}+(\frac{15}{2}w_x-\frac{105}{2}v_x)v_{xxx}-45v_{xx}^2+\frac{45}{2}pv_xv_{xx}+15pv_{xy}+45w_{xx}^2\\
	&+(\frac{225}{2}pw_x-45pv_x+\frac{45}{4}p^3)w_{xx}+\frac{45}{4}v_x^2p_x+\frac{135}{4}p^2w_xp_x+15p_xv_y+\frac{135}{4}w_x^2p_x+\frac{45}{16}p^4p_x,
	\end{aligned}\right.
\end{equation}
in which case, the decomposition solution depends on another solution $v$ of the potential BKP equation \eqref{pbkp}. 
	
	\bf \it Case 3. $H_1=-\frac12$. \rm In this case, the functions $H$ and $J$ are fixed as
	\begin{equation}\label{HJ_KdV3}
	\left\{
	\begin{aligned}
	H&=-\frac{3}{2}w_x^2+6Mw_x+\frac{y}{10}(m_1y+2m_2)_t+m_1x-6M^2+m_3, M\equiv m_1y+m_2,\\
	 J&=-\frac{45}{4}(w_{xx}^2+2w_x^3)+15m_1w+\frac{3}{2}y(m_1y+2m_2)_tw_x+15(6M^2+m_1x+m_3)w_x+\frac{y^2}{30}(m_1y\\&+3m_2)_{tt}-3(y^2m_{1t}+10m_1x+2ym_{2t})M+M_tx+(m_{3t}-30m_1m_3)y-60M^3+m_4,
	\end{aligned}\right.\end{equation}
		where $m_i, i = 1, 2, 3, 4$ are arbitrary functions of $t$.

	By substituting these solutions \eqref{HJ_SK}--\eqref{HJ_KdV3} into the decomposition relations \eqref{wy1} and \eqref{wt1}, Theorems 1-5 are proved.
\end{proof}

\end{document}